\documentclass[a4paper,11pt]{article}

\usepackage[tbtags]{amsmath}
\usepackage{amssymb}
\usepackage{graphicx}
\usepackage{times}
\usepackage{wrapfig}
\usepackage[small,center]{caption}
\usepackage[colorlinks=false]{hyperref}

\newcommand\Imag{\text{Im}} 

\providecommand\bnabla{\boldsymbol{\nabla}}
\newcommand\ud{\mathrm{d}}
\newcommand\bR{\mathbf{R}}
\newcommand\bS{\mathbf{S}}

\newcommand\biX{\boldsymbol{X}}
\newcommand\bia{\boldsymbol{a}}
\newcommand\biO{\boldsymbol{\varOmega}}
\newcommand\biS{\boldsymbol{S}}

\newcommand\const{\text{const}}

\begin{document}

\DeclareGraphicsExtensions{.eps}

\begin{center}
    {\large \textbf{MATRIX FLUID DYNAMICS}}\\
\vspace{12pt}
     \textbf{E. I. Yakubovich}\; and\; \textbf{D. A. Zenkovich} \\
\vspace{6pt}
   {\small Institute of Applied Physics, Nizhny Novgorod, Russia}  
\end{center}

The lagrangian representation of fluid motion is alternative with respect to the conventional eulerian description. It is focused on observing the motion of individual fluid particles
identified by three parameters known as the lagrangian variables. The set of ideal-fluid lagrangian governing equations consists of four non-linear differential equations for current positions of fluid particles as functions of the lagrangian variables and time~\cite{lamb}. According to Aref~(\cite{aref}, p.274), `\dots analytical tradition has clearly favoured the eulerian representation, and considered the lagrangian version to be analytically intractable'. The lagrangian analytical study of a viscous fluid is even more problematic because of the extreme complexity of the known expression for the viscous force~\cite{monin-yaglom}.

This paper aims at showing that we can advance in the analytical study of the lagrangian fluid dynamics due to a new matrix approach proposed in~\cite{mat-jfm}. The matrix approach is  based on the notion of continuous deformation of infinitesimal vector fluid elements and treats the Jacobi matrix of derivatives of particles coordinates with respect to the lagrangian variables as the fundamental quantity describing fluid motion. The matrix (`deformation') approach is distinct from the conventional lagrangian (`trajectory') formulation and is, actually, similar to the tensor description in the elasticity theory. A closed set of governing equations  is formulated in terms of the Jacobi matrix and studied for both ideal and viscous fluid. Recent results on the 3-D extension of plane lagrangian solutions are reported.

\vspace{12pt}
\textbf{Matrix form of ideal-fluid hydrodynamic equations}
\vspace{6pt}

In the conventional lagrangian approach, fluid motion is described
by current particle position $\biX= \{ X, Y,Z \}$, considered as a function of time $t$ and the lagrangian variables $\bia=\{a, b, c\}$: $\biX=\biX(\bia,t)$. Let us consider the Jacobi matrix $\bR (\bia,t)=\partial \biX / \partial \bia$ as the principal characteristic of fluid motion. The evolution of an infinitesimal material vector $\ud \biX$ corresponding to an increment of the lagrangian variables $\ud \bia$ is then given by $\ud \biX= \bR \, \ud \bia$. For an ideal fluid, the Jacobi matrix is shown to satisfy the following set of equations~\cite{mat-jfm}:
\begin{equation}\label{set1}
  \det \bR=\det \bR_0, \quad 
  \bR_t^{\mathrm{T}} \bR - \bR^{\mathrm{T}} \bR_t =\bS, \quad \partial R_{nm}/\partial a_k=\partial R_{nk}/\partial a_m,
\end{equation}
where $\bR_0=\bR(\bia,0)$, $\bR^{\mathrm{T}}$ is the transposed Jacobi matrix, subscript `$t$' denoting the time derivative,  $\bS = \left( {e_{ijk} S^k } \right)$ is a time-independent antisymmetric matrix of the Cauchy invariants, $\{ a,b,c\} \equiv \{ a_1 ,a_2 ,a_3 \}$ for the indexed notation.  The Cauchy invariants $S^k$ are the integrals of the conventional 3-D lagrangian equations related to the initial vorticity $\biO _{0}$ at $t=0$ by $S^k(\bia) = ( \biO _{0} ,\bnabla a_k )\det \bR_0 $. Set (\ref{set1}) comprises the matrix analogues of the lagrangian continuity  and vorticity equations along with the consistency condition. This condition allows us to reconstruct the particle trajectories $\biX(\bia, t)$  by a solution of (\ref{set1}) for $\bR$ uniquely through integration over the lagrangian variables:
\begin{equation}\label{eq4}
\textstyle  \biX(\bia,t) = \int_{0}^{\bia} {\bR(\boldsymbol{q}, t)\, \ud \boldsymbol{q}}.
\end{equation}
 
The specific features of set \eqref{set1} consist in the following. The continuity equation takes the form of an algebraic constraint on the Jacobi matrix. The equation of motion is formulated as a nonlinear matrix differential equation in time only, where the derivatives with respect to the lagrangian variables do not appear.  The accompanying consistency condition is linear and does not include the time derivatives. Altogether, set \eqref{set1} admits a flexible formulation of the problem of interest and allows one to take advantage of the
powerful machinery of matrix calculus. 
 
The expression for vorticity takes the form
\begin{equation}\label{e5}
\biO  = (\det \bR_0)^{-1} S^i \;\partial \biX / \partial a_i 
\end{equation}
and represents decomposition of vorticity into components along the reference vectors $\partial \biX / \partial a_i$ of the frozen-in `liquid' coordinate frame of the lagrangian variables. Accordingly, the meaning of the invariants $S^i$ is that they are proportional to the contravariant components of vorticity in the `lagrangian' reference frame. As seen from the formula for its Cartesian components $\varOmega_n  =  \det \bR_0^{-1} R_{ni} \,S^i$, the evolution of vorticity is described by the Jacobi matrix directly.

Thus, by virtue of the facts that set \eqref{set1} is closed and one can reconstruct any required characteristic of the flow by its solution for $\bR$ it follows that the matrix approach based on set \eqref{set1} provides a complete description of motion of an ideal fluid. 

A method for constructing solutions to the set of matrix equations which correspond to 3-D rotational flows with precessing vorticity is developed in \cite{mat-jfm}.  It allowed us to find and investigate a new class of 3-D exact matrix solutions that would be extremely difficult to derive from conventional formulations. The solutions investigated in \cite{mat-jfm} depend not only on a set of parameters but on several arbitrary functions of the Lagrangian variables and incorporate the known 2-D solutions such as the Ptolemaic vortices \cite{dan}, the Gerstner waves and Kirchhoff's vortex as particular cases. Below we report  results of an alternative technique to construct 3-D solutions of the matrix equations \eqref{set1}.


\vspace{12pt}
\textbf{3-D shearing and stretching of plane motion}
\vspace{6pt}

The matrix formulation provides a compact `block' representation of 3-D lag\-ran\-gian equations and allows us to study new 3-D extensions of known 2-D rotational motions. Suppose that $\widetilde{X}(a,b,t), \widetilde{Y}(a,b,t)$ represent a solution of 2-D lagrangian hydrodynamic equations. The corresponding Jacobi matrix takes the form
\begin{equation}\label{e21}
\widetilde \bR = \left[ {\begin{array}{ccc}
   {\widetilde{X}_a (a,b,t)} & {\widetilde{X}_b (a,b,t)} & 0  \\
   {\widetilde{Y}_a (a,b,t)} & {\widetilde{Y}_b (a,b,t)} & 0  \\
   0 & 0 & 1  \\
 \end{array} } \right].
\end{equation}
It can be proved that along with matrix (\ref{e21}), set (\ref{set1}) is satisfied by the transformed matrix 
\begin{equation}\label{e22}
\mathbf{R} = \left[ {\begin{array}{ccc}
   f^{-1/2} \widetilde{X}_a \left(a,b,\int {f\ud t} \right) & f^{-1/2} \widetilde{X}_b \left(a,b,\int {f \ud t} \right) & 0  \\
f^{-1/2}\, \widetilde{Y}_a \left(a,b,\int {f \ud t} \right) & f^{-1/2}\, \widetilde{Y}_b\left(a,b,\int {f\ud t} \right) & 0  \\
   f \left(\int {f^{ - 2} \ud t}\right)H_a (a,b) &
   f \left(\int {f^{ - 2} \ud t}\right) H_b (a,b) & f(t)  \\
 \end{array} } \right],
\end{equation}
which includes an arbitrary function of the lagrangian variables $H(a, b)$ and an arbitrary function of time $f(t)\neq 0$.  The particle trajectories obtained from (\ref{e22}) using (\ref{eq4}) take the form
\begin{equation}\label{e23}
  X = \frac{ \widetilde{X}\left(a,b, \int {f\ud t} \right)}{f^{1/2}},\;\; Y = \frac{\widetilde{Y}\left(a,b,\int {f\ud t} \right) }{f^{1/2}} , \;\;
 \textstyle Z = f \left[H(a,b) \int f^{ - 2} \ud t + c \right]
\end{equation}
and represent a 3-D motion with the $Z$-component of velocity
 \begin{equation} \label{e24} V_z  = [H(a,b) + \dot f Z]/f. \end{equation}
The Cauchy invariants of the 3-D solution are given by
\begin{equation}\label{e25}
S^1  = H_b ,\quad S^2  =  - H_a ,\quad S^3  = \widetilde X_{ta} \widetilde X_b  - \widetilde X_{tb} \widetilde X_a  + \widetilde Y_{ta} \widetilde Y_b  - \widetilde Y_{tb} \widetilde Y_a ,
\end{equation}
where $S^3$ remains the same as for the 2-D solution (\ref{e21}), but both $S^1$ and $S^2$ are now non-zero. The vorticity corresponding to (\ref{e22}), (\ref{e23}) is obtained from (\ref{e5}) using (\ref{e25}), it becomes 3-D and time-dependent:
\begin{equation}\label{e26}
\biO  = \left\{ {f^{ - 1} \partial H/\partial Y,\; - f^{ - 1} \partial H/\partial X,\;S^3 f/(\widetilde X_a \widetilde Y_b  - \widetilde X_b \widetilde Y_a )} \right\}.
\end{equation}

The transformation according to (\ref{e22}), (\ref{e23}) allows one to obtain non-trivial extensions of any given 2-D solution. As can be seen from (\ref{e24}), (\ref{e26}), the contributions of arbitrary functions $H(a, b)$ and $f(t)$ to the resulting 3-D motion are a shear flow along the $Z$-direction and stretching of the $Z$-component of vorticity, respectively. We shall illustrate the effect of these two factors separately using the 2-D Ptolemaic vortices \cite{mat-jfm,dan} as the initial solution. The lagrangian expression for the flow inside the 2-D Ptolemaic vortex is given in complex variables \cite{dan}:
\begin{equation}\label{e27}
W  = 2^{- 1/2} (X + iY) = \xi \exp( - 2i\omega t)  + F(\overline \xi )\quad \text{for} \quad |\xi| < 1,
\end{equation}
where $W$ is the complex coordinate, $\xi = 2^{-1/2}(a+ib)$ is the complex lagrangian variable, $\overline \xi$ denotes the complex conjugate variable, function $F(\overline \xi)$ is analytical within $|\xi| < 1$ and satisfies there the condition $|F'| < 1$.

\vspace{6pt}
\underline{\textit{3-D Ptolemaic vortices with swirl}}
\vspace{0pt}

In order to focus on the effect of the $Z$-shear, we apply transformation (\ref{e23}) to (\ref{e27}) and eliminate stretching by setting $f(t)=1$. The resulting flow is most expediently described in complex form similar to the initial solution \eqref{e27}:
\begin{equation}\label{e28}
W  = \xi \exp ( - 2i\omega t) + F(\overline \xi),\quad Z = H(\xi,\overline \xi )\,t + c.
\end{equation}
Here, $H(\xi, \overline \xi)$ remains real-valued. The particle trajectories are regular circular helices. Although the lagrangian expression for the $Z$-component of the flow is elementary, the eulerian velocity for (\ref{e28}) cannot be written explicitly and is evidently non-trivial. The Cartesian components of vorticity are derived from (\ref{e26}), (\ref{e28}):
\begin{equation}\label{e29}
\varOmega _X  + i\varOmega _Y  = \frac{ - \sqrt 2 \,i}{1 - |F'|^2 }\left[ H_{\overline \xi } e^{ - 2i\omega t}  - H_{\xi}  e^{2i\omega t} F'(\overline \xi ) \right],\quad \varOmega _Z  = \frac{ - 4\;\omega }{1 - |F'|^2 }.
\end{equation}
\begin{wrapfigure}{R}{4.2cm}
\includegraphics[height=6cm]{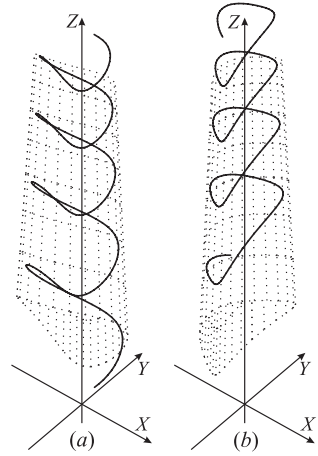}
\caption{Vortex lines (solid) and vortex tubes (dotted) for (\ref{e28}) (see the above text for detail).}
\label{fig4}
\end{wrapfigure}
To have a geometrical idea of the structure of vortex lines, consider their equation in the space of the lagrangian variables $\ud a/S^1  = \ud b/S^2  = \ud c/S^3$. The substitution of the Cauchy invariants $S^i$ from (\ref{e25}) yields $\ud a/H_b = - \ud b/H_a =  - \ud c /(4\omega)$, then $H=\const$ along the vortex lines. If the contours of $H(a, b)=\const$ are closed inside the vortex (for $|a+ib|<1$), the vortex lines are windings on the cylindrical surfaces in the space of $a, b, c$, for which these contours serve as directing curves. The shapes of vortex lines in Cartesian space are topologically similar, although they experience  periodic distortions, since the smooth mapping defined by (\ref{e28}) is  time-dependent. Figure~\ref{fig4} shows the material (moving with fluid particles) vortex lines of flow (\ref{e28}) for $|\xi|=0.9$, and the surface of the vortex tube for $|\xi|=1$: (\textit{a}) at  $t = 0$, (\textit{b})  at $t = 2\pi$; here  $F = \alpha (\overline \xi  + \beta )^{5/2}$,\, $\alpha = 0.11$,\, $\beta = 1.2$,\, $H = 2(1 - |\xi|^2 )$,\, $\omega  = 1/4$.

Such an important solution of 2-D lagrangian equations as the Gerstner waves \cite{lamb,dan}, can be extended by analogy with (\ref{e28}):
\begin{equation}\label{e30}
W  = \xi  + A\exp [i(k\overline \xi  + \sigma t)],\;\; Z = H(\xi ,\overline \xi) t + c.
\end{equation}
Noteworthy is the fact that (\ref{e30}) satisfies not only the governing equations but also the boundary conditions at the free surface $\Imag\, \xi =0$.
The resulting 3-D flows have helical particle trajectories and generally curvilinear vortex lines, their shape depending on the real-valued function $H(\xi ,\overline \xi)$. 

\vspace{6pt}
\underline{\textit{Stretched Ptolemaic vortices}}
\vspace{0pt}

Assume that the $Z$-shear in (\ref{e22})--(\ref{e24}) vanishes with $H=0$, and consider the effect of the arbitrary function $f(t)$ proceeding from 2-D Ptolemaic vortices (\ref{e27}). We shall demonstrate that the extension according to (\ref{e23}) leads to stretching or contraction of the vortex by a non-stationary axisymmetric strain flow. The particle trajectories obtained from (\ref{e27}) for $H=0$, $f(t) \neq 0$ take the form
\begin{equation}\label{e31}
W  = f^{ - 1/2} \left[ \xi \exp\left(  - 2i\omega \textstyle \int f\,\ud t  \right) + F(\overline \xi ) \right],\;\; Z = f(t)c.
\end{equation}
For monotonic $f(t)$, all the particles trace convergent\,/\,divergent spiral paths.
Solution (\ref{e31}) has one non-zero Cauchy invariant $S^3=- 4\,\omega$ and the only component of vorticity $\varOmega_Z=- 4\omega f(t) /\left( {1 - |F'|^2 } \right)$ which depends on the lagrangian variables and time. It is the time variation of vorticity in proportion to $f(t)$ that reflects the effect of concentration and intensification of vorticity (for $\ud |f| / \ud t >0$) due to the stretching of vortex tubes along the $Z$-direction. 

The manifestation of the strain flow becomes evident for the isolated vortex (\ref{e31}) streamlined by an irrotational fluid. We now assume that the 3-D rotational motion (\ref{e31}) is concentrated within a vortex core, e.g. within the vortex tube $|\xi|\leqslant 1$.  It proves possible due to a specially developed technique of potential continuation to construct an exterior 3-D potential flow which agrees with the rotational motion across the boundary of the vortex. We formulate a system of parametric matrix equations for the exterior potential velocity, provided that parameterization of the coordinates in the outer region is given. The set of parameters for the outer potential region is most expediently chosen in accordance with the parameterization of the interior by the lagrangian variables (\ref{e31}). Using the matrix equations of potentiality we obtain the potential velocity parametrically at the point of the outer region
\begin{equation}\label{e32}
W  = f^{- 1/2} \left[ \xi \exp \left(  - 2i\omega \textstyle \int {f\ud t}  \right) + F(1/\xi) \right],\; Z = f(t)c \quad \text{for} \quad |\xi| > 1
\end{equation}
as a function of parameters $\xi, c$\,:
\begin{equation}\label{e33}
V  =  2^{-1/2} (V_x  + iV_y ) =  - 2i\omega f^{1/2}  e^ {- 2i\omega \int {f\ud t}} / \, \overline \xi - \textstyle \frac{1}{2}\, W\,\dot f /f ,\quad V_Z  = \dot f c,
\end{equation}
where $\dot f$ stands for the time derivative of $f$.
\begin{wrapfigure}[15]{R}{5cm}
\includegraphics[height=5cm]{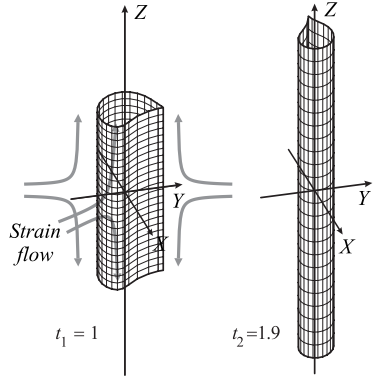}
\caption{Stretching of a material section of the vortex \eqref{e31}--\eqref{e33}\newline in a strain flow.}
\label{fig5}
\end{wrapfigure}
All the components of (\ref{e33}) agree with the velocity obtained from (\ref{e31}) across the vortex boundary $|\xi| = 1$.
Unlike the vortex core,  for the outer region $\xi, c$ are not the lagrangian variables of
the potential flow and play the role of formal parameters in (\ref{e32}), (\ref{e33}).
As the distance from the vortex increases ($|\xi| \rightarrow \infty$), the potential velocity (\ref{e33}) tends asymptotically to the non-stationary axisymmetric strain flow, the intensity of which is proportional to $\dot f$: $V_r  \asymp  - r\dot f/(2f) + O(1/r)$, $V_z  = Z \dot f/f$, where $r$ is the cylindrical radius. It becomes clear that the stretching or contraction of the vortex is influenced by the external strain and is essentially kinematical.
Figure \ref{fig5} shows the boundary of the material section $|\xi| = 1$, $|c|\leqslant 4$ of the vortex (\ref{e31})--(\ref{e33}) for $f(t) = t$ and $F(\overline \xi) = \alpha \left[ (\overline \xi  - \zeta_1 )^{ - 3}  + (\overline \xi  - \zeta_2)^{-3} \right]$, $\alpha = 0.83$, $\zeta_1=2.2$, $\zeta_1=2.09+0.68i$. 

We have demonstrated the effects of 3-D shear and axisymmetric strain on 2-D lagrangian solutions in succession. As seen from general expressions \eqref{e22}, (\ref{e23}), the extension of 2-D solutions allowing for their concurrent contribution is also simple and results in a variety of interesting 3-D flows.

\vspace{12pt}
\textbf{Lagrangian equations for a viscous fluid}
\vspace{6pt}

The notion of the frozen-in lagrangian coordinates and the matrix formulation are useful not only for an ideal fluid but also for a viscous fluid. We have seen that the constancy of the Cauchy invariants $S^i$ plays an important role for an ideal fluid. It is natural to question how they change due to viscous force. We shall derive an equation for $S^i$ which serves as the lagrangian analogue of the Helmholtz equation for a viscous fluid.

The lagrangian equations deduced by a direct change to the lagrangian variables in the Navier--Stokes equation are known from the literature \cite{monin-yaglom}. However, their structure is so complicated that no progress in their analytical study or applications has been achieved until now.

For an alternative approach, we consider the column vector $\biS = \{S^1, S^2, S^3\}$ composed of the Cauchy invariants. It is related to the velocity vector by $\biS= \mathrm{rot}_{\bia}(\bR^\mathrm{T} \biX_t)$, where $(\mathrm{rot}_{\bia} \boldsymbol{A})_i=e_{ijk} \partial A_j / \partial a_k$. The change of $\biS$ occurs due to the viscous force exclusively and is governed by the diffusion-type equation
\begin{equation}\label{e34}
\partial \biS / \partial t= - \eta \,\mathrm{rot}_{\bia} \left[ \mathbf{g} \; \mathrm{rot}_{\bia} \left( \mathbf{g}\,\biS /  D  \right) / D \right],
\end{equation}
where $\eta$ is the kinematic viscosity,  $\mathbf{g}=(g_{ik})=\bR^{\mathrm{T}}\bR$ is the metric tensor reflecting the distortion of the frozen-in `liquid' coordinate system, $D=(\det \mathbf g)^{1/2}$. This equation complements set \eqref{set1} of the ideal-fluid matrix equations. Together, they constitute a closed system of equations that fully describes the motion of a viscous fluid, provided that $S^i$ in \eqref{set1} are considered as functions of time.

For a 2-D motion $S^1=S^2=0$ and \eqref{e34} reduces to 
\begin{equation}\label{e35}
\frac{{\partial S}}
{{\partial t}} = \eta \left\{ {\partial _a  \left[ {\frac{{g_{21} }}
{D}\partial _b  \left( \negthinspace \frac{S}
{D} \right) - \frac{{g_{22} }}
{D}\partial _a  \left( \negthinspace \frac{S}
{D} \right)} \right] - \partial _b  \left[ {\frac{g_{11}}
{D}\partial _b  \left(\negthinspace \frac{S}
{D} \right) - \frac{{g_{12} }}
{D}\partial _a  \left(  \negthinspace \frac{S}
{D} \right)} \right]} \right\},
\end{equation}
where $S \equiv S^3$. Equation \eqref{e35} can be simplified asymptotically for low or high viscosity. For the case of low viscosity ($\mathit{Re} \ll 1$), the tensor components $g_{ik}$ can be approximately replaced by their time average obtained from the inviscid problem. For high viscosity ($\mathit{Re} \gg 1$), the term $\partial S / \partial t$ on the left-hand side of \eqref{e35} can asymptotically be neglected, so that time plays the role of a parameter in \eqref{e35} (note that we should nevertheless keep the time derivatives in \eqref{set1}). More profound study of \eqref{e34}, \eqref{e35} and their applications are beyond the scope of this report.

The authors gratefully acknowledge the financial
support provided by INTAS Projects 97-0575, 99-1637 and Project
No.\,00-15-96772 of the Russian Foundation for Basic Research.

\end{document}